\documentclass[%
    reprint,
    aps,
    prl,
    superscriptaddress,
    floatfix,
    bibnotes,
    amsmath,
    amssymb,
    amsfonts,
]{revtex4-2}

\usepackage[utf8]{inputenc}
\usepackage[T1]{fontenc}
\usepackage[svgnames]{xcolor}
\usepackage{dcolumn}
\usepackage{bm}
\usepackage{enumerate}
\usepackage{enumitem}
\usepackage{csquotes}
\usepackage{empheq}
\usepackage{dsfont}
\usepackage{siunitx}
\usepackage{physics}
\usepackage{cancel}
\usepackage{hyperref}
\usepackage[caption=false]{subfig}
\usepackage{graphicx}
\usepackage{cleveref}

\hypersetup{
	colorlinks  = true,
	citecolor   = Orange!80!black,
	linkcolor   = RoyalBlue!60!black,
	urlcolor    = RoyalBlue!80!black,
	unicode     = true,
	pdfencoding = auto
}

\makeatletter
\newcommand{\beginsupplement}{%
  \clearpage
  \onecolumngrid
  \@removefromreset{equation}{section}%
  \setcounter{section}{0}%
  \setcounter{subsection}{0}%
  \setcounter{subsubsection}{0}%
  \setcounter{equation}{0}%
  \setcounter{figure}{0}%
  \setcounter{table}{0}%
  \setcounter{page}{1}%
  \renewcommand{\thepage}{S\arabic{page}}%
  \renewcommand{\thesection}{S\arabic{section}}%
  \renewcommand{\thesubsection}{\thesection.\arabic{subsection}}%
  \renewcommand{\thesubsubsection}{\thesubsection.\arabic{subsubsection}}%
  \renewcommand{\theequation}{S\arabic{equation}}%
  \renewcommand{\thefigure}{S\arabic{figure}}%
  \renewcommand{\thetable}{S\arabic{table}}%
  \renewcommand{\theHsection}{supp.\arabic{section}}%
  \renewcommand{\theHsubsection}{supp.\arabic{section}.\arabic{subsection}}%
  \renewcommand{\theHsubsubsection}{supp.\arabic{section}.\arabic{subsection}.\arabic{subsubsection}}%
  \renewcommand{\theHequation}{supp.\arabic{equation}}%
  \renewcommand{\theHfigure}{supp.\arabic{figure}}%
  \renewcommand{\theHtable}{supp.\arabic{table}}%
}
\makeatother

\usepackage{tikz}
\usetikzlibrary{arrows.meta, positioning, fit, backgrounds}
\usepackage{pgfplots}
\pgfplotsset{compat=1.18}
\usepgfplotslibrary{fillbetween}

\tikzstyle{site} = [circle, shading=ball, ball color=black!40, minimum size=4mm, inner sep=0pt]

\renewcommand{\vec}[1]{\ensuremath{\bm{#1}}}

\begin{document}

\title{Universal Symmetry-Breaking Dynamics at Continuous Phase Transitions: Evidence for a New Dynamical Critical Exponent}

\author{Tobias \surname{Wiener}}
\email{tobias.wiener@uni-a.de}
\affiliation{Theoretical Physics III, Center for Electronic Correlations and Magnetism, Institute of Physics, University of Augsburg, D-86135 Augsburg, Germany}

\author{Laurin \surname{Brunner}}
\email{laurin.brunner@uni-a.de}
\affiliation{Theoretical Physics III, Center for Electronic Correlations and Magnetism, Institute of Physics, University of Augsburg, D-86135 Augsburg, Germany}

\author{Markus \surname{Heyl}}
\email{markus.heyl@uni-a.de}
\affiliation{Theoretical Physics III, Center for Electronic Correlations and Magnetism, Institute of Physics, University of Augsburg, D-86135 Augsburg, Germany}

\date{\today}

\begin{abstract}
Uncovering and understanding universal dynamics in matter far from equilibrium remains a key challenge. In this work, we identify a so far unrecognized form of universal behavior that emerges after a sudden symmetry-breaking quench at continuous phase transitions. Our key observation is that the order-parameter fluctuations in Ising models exhibit a compelling temporal collapse across a wide range of system sizes and quench strengths, indicative of an emergent single-variable scaling form. This phenomenon can be explained by introducing a so far unknown dynamical critical exponent for the underlying continuous phase transition. We find evidence for a lower critical effective dimension of this universal regime: it is observed in the 2D quantum and 3D and 4D classical Ising models, but not in the 1D quantum or 2D classical cases. Our results suggest that our observed universal far-from-equilibrium scaling may extend beyond the Ising models studied here and could more broadly characterize systems with non-conserved order parameters, opening new avenues for exploring universal dynamics both theoretically and in current experimental platforms.
\end{abstract}

\maketitle

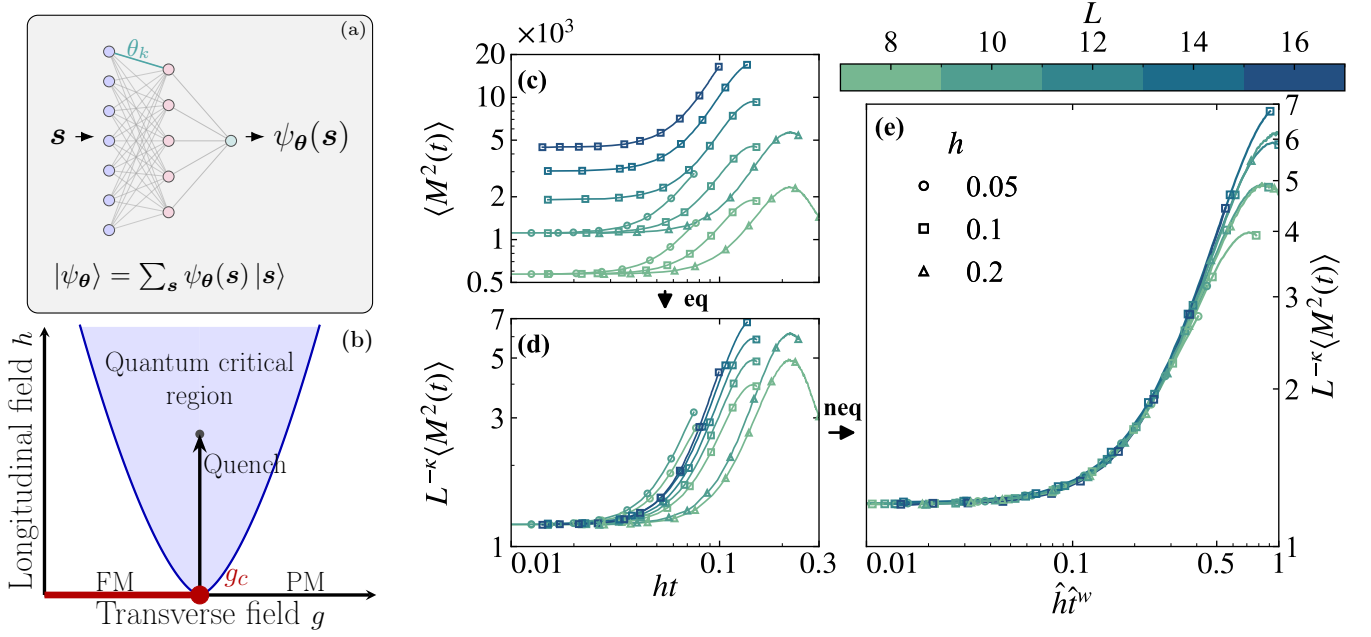
\begin{figure*}[t]
    \centering
    \newsavebox{\figcollapseleft}
    \begin{lrbox}{\figcollapseleft}
        \begin{minipage}[b]{0.28\textwidth}
            \centering
            \resizebox{\linewidth}{!}{\begingroup

\newcommand{\nnicon}{%
\tikz[baseline=-0.6ex, scale=0.78]{
  \tikzset{
    nnode/.style={circle, draw=black!70, line width=0.25pt, inner sep=0pt, minimum size=4.1pt},
    inlayer/.style ={nnode, fill=blue!18},
    hid/.style ={nnode, fill=purple!16},
    outlayer/.style ={nnode, fill=teal!18},
    conn/.style={line width=0.35pt, black!35, line cap=round},
    connparam/.style={line width=0.6pt, teal!70}
  }

  \def\xI{0.0} \def\xH{1.0} \def\xO{2.05}

  \foreach \y/\n in {1.5/i1,1.0/i2,0.5/i3,0/i4,-0.5/i5,-1.0/i6,-1.5/i7}
    \node[inlayer] (\n) at (\xI,\y) {};

  \foreach \y/\n in {1.2/h1,0.6/h2,0/h3,-0.6/h4,-1.2/h5}
    \node[hid] (\n) at (\xH,\y) {};

  \node[outlayer] (o1) at (\xO, 0) {};

  \foreach \i in {1,...,7}{
    \foreach \h in {1,...,5}{
      \draw[conn, opacity=0.55] (i\i) -- (h\h);
    }
  }
  \draw[connparam] (i1) -- (h1) node[midway, above, font=\small, inner sep=0.5pt] {$\theta_k$};
  \foreach \h in {1,...,5}{
    \draw[conn, opacity=0.7] (h\h) -- (o1);
  }
}}
\newlength{\boxw}
\setlength{\boxw}{1.0\linewidth}

\begin{tikzpicture}[
  font=\large,
  node distance=2.2mm and 2.5mm,
  box/.style={
    draw,
    fill=gray!10,
    rounded corners,
    align=center,
    inner sep=5pt,
    minimum height=8.0mm
  },
  arr/.style={-Latex, line width=0.45pt},
  quench/.style={draw=red!75!black},
  quencharr/.style={-Latex, line width=0.65pt, red!75!black},
]

\node (s) {$\vec{s}$};
\node[right=of s] (nn) {\nnicon};
\node[right=of nn] (eq) {$\psi_{\vec{\theta}}(\vec{s})$};
\draw[arr] (s) -- (nn);
\draw[arr] (nn) -- (eq);

\node[below=1.2mm of nn, font=\normalsize, align=center] (eqline) {$\ket{\psi_{\vec{\theta}}} = \sum_{\vec{s}} \psi_{\vec{\theta}}(\vec{s}) \ket{\vec{s}}$};

\scoped[on background layer]{%
  \node[fit=(s)(nn)(eq)(eqline), draw, fill=gray!10, rounded corners, inner sep=5pt] (init) {};
}

\node[anchor=north east, font=\scriptsize, inner sep=0pt]
  at ([xshift=-2pt,yshift=-2pt]init.north east) {(a)};

\end{tikzpicture}

\endgroup}
            \\[0.3em]
            \resizebox{\linewidth}{!}{\usetikzlibrary{arrows.meta}
\begin{tikzpicture}
    \begin{axis}[
        width=12cm,
        height=10cm,
        xlabel={Transverse field $g$},
        ylabel={Longitudinal field $h$},
        xlabel style={font=\fontsize{28}{34}\selectfont},
        ylabel style={font=\fontsize{28}{34}\selectfont},
        xmin=0.78, xmax=1.25,
        ymin=0, ymax=0.05,
        axis lines=left,
        axis line style={line width=3pt},
        xtick=\empty,
        ytick=\empty,
        legend style={at={(0.97,0.97)},anchor=north east,font=\Huge},
        clip=false
    ]
    
    \def\nuval{0.63}
    \def\pval{1.56}
    
    \addplot[
        domain=0.83:1,
        samples=100,
        name path=left,
        color=blue!70!black,
        thick,
        forget plot
    ] {0.8*(1-x)^\pval};
    
    \addplot[
        domain=1:1.17,
        samples=100,
        name path=right,
        color=blue!70!black,
        thick,
        forget plot
    ] {0.8*(x-1)^\pval};
    
    \path[name path=top] (axis cs:0.83,0.05) -- (axis cs:1.17,0.05);
    
    \addplot[
        blue!20,
        opacity=0.6
    ] fill between[
        of=left and top,
        soft clip={domain=0.83:1}
    ];
    
    \addplot[
        blue!20,
        opacity=0.6
    ] fill between[
        of=right and top,
        soft clip={domain=1:1.17}
    ];
    
    \addplot[
        domain=0.83:1,
        samples=100,
        color=blue!70!black,
        thick,
        line width=2pt
    ] {0.8*(1-x)^\pval};
    
    \addplot[
        domain=1:1.17,
        samples=100,
        color=blue!70!black,
        thick,
        line width=2pt
    ] {0.8*(x-1)^\pval};
    
    \addplot[
        mark=*,
        mark size=8pt,
        color=red!70!black,
        only marks
    ] coordinates {(1,0)};
    
    \draw[red!70!black, line width=6pt] (axis cs:0.78,0) -- (axis cs:1,0);

    \node[font=\Huge, anchor=south] at (axis cs:0.88,0) {FM};
    \node[font=\Huge, anchor=south] at (axis cs:1.15,0) {PM};
    
    \node[font=\Huge,color=red!70!black,anchor=south west] at (axis cs:1.03,0.001) {$g_c$};
    
    \fill[gray!60!black] (axis cs:1,0.03) circle (4pt);
    
    \draw[-Stealth,thick,black,line width=3pt]
    (axis cs:1,0) -- (axis cs:1,0.03)
    node[pos=0.8, right, inner sep=1.5pt, font=\Huge] {Quench};
  
    \node[font=\Huge, anchor=center, align=center, inner sep=2pt]
      at (axis cs:1,0.040)
      {Quantum critical\\region};
    
    \coordinate (topright) at (axis cs:1.25,0.05);
    \node[anchor=north east, font=\fontsize{22}{26}\selectfont\bfseries, inner sep=0pt]
      at ([xshift=-2pt,yshift=-2pt]topright) {(b)};
    \end{axis}
\end{tikzpicture}}
        \end{minipage}
    \end{lrbox}
    \usebox{\figcollapseleft}\hfill
    \begin{minipage}[b]{0.70\textwidth}
        \centering
        \includegraphics[width=\linewidth, height=\dimexpr\ht\figcollapseleft+\dp\figcollapseleft+10em\relax, keepaspectratio]{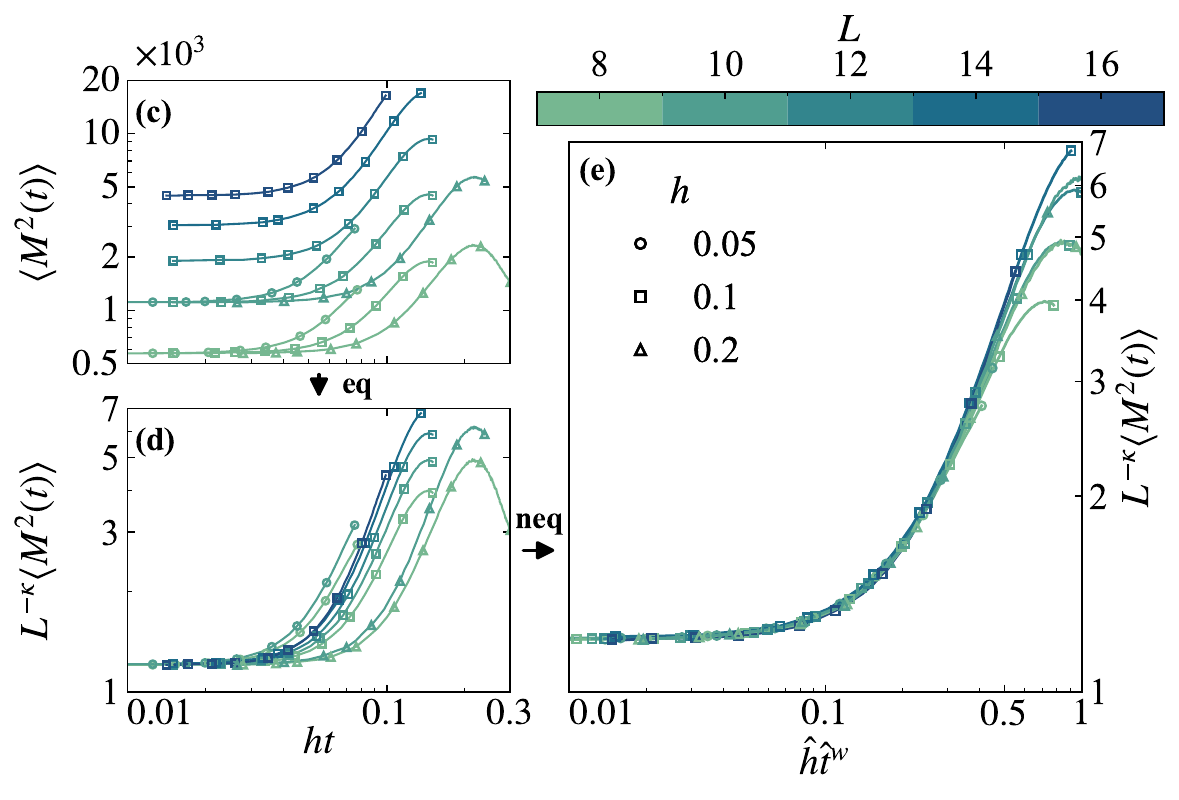}
    \end{minipage}
    \caption{
        \textbf{Symmetry-breaking quench at the 2D TFIM critical point and emergent single-variable scaling.}
        (a) Neural Quantum State (NQS) ansatz: the many-body wave function is represented by a neural network, $\ket{\psi_{\vec{\theta}}}=\sum_{\vec{s}}\psi_{\vec{\theta}}(\vec{s})\ket{\vec{s}}$.
        (b) Ground-state phase diagram in the $(g,h)$ plane and quench path from $(g_c,0)$ to $h\neq 0$ at fixed $g=g_c$.
        (c) Raw $\langle M^2(t)\rangle$ versus $ht$ for various $L$ and $h$.
        (d) Same data after equilibrium rescaling $L^{-\kappa}\langle M^2(t)\rangle$ versus $ht$.
        (e) Full collapse of the same rescaled observable versus $\hat h \hat t^w$, where $\hat t=(Jt)L^{-z}$ and $\hat h=(h/J)L^{y_h}$; the best collapse is obtained for $w=1.86\pm0.11$.
        Colors denote $L$ and symbols denote $h$.
    }
    \label{fig:collapse}
\end{figure*}

\paragraph{Introduction.}
Universality is one of the central organizing principles of many-body physics: near continuous phase transitions, microscopically distinct systems can exhibit identical behavior characterized by a small set of universal exponents. This concept extends beyond statics into the dynamical regime. Close to equilibrium, relaxation processes are systematically classified within the Hohenberg--Halperin framework~\cite{hohenberg_theory_1977,tauber_critical_2014,pelissetto_dynamic_2018}. Far from equilibrium, universal scaling has so far been established mainly for specific protocol classes, including the Kibble--Zurek mechanism for slow ramps across a transition~\cite{kibble_topology_1976,zurek_cosmological_1985,zurek_cosmological_1996,zurek_dynamics_2005,delcampo_universality_2014,keesling_quantum_2019}, initial-slip and short-time scaling following sudden quenches from a disordered state to criticality~\cite{janssen_new_1989,zheng_generalized_1996,okano_shorttime_1997,zheng_monte_1998,rossini_coherent_2021,chiocchetta_shorttime_2015,chiocchetta_lightcone_2016}, and non-thermal fixed points after deep quenches in isolated systems~\cite{schmied_nonthermal_2019}. However, dynamical universal behavior far from equilibrium, particularly when fluctuations dominate and push the system beyond any linear-response regime, remains largely an open question. This is especially true for sudden symmetry-breaking quenches \emph{at} a continuous phase transition, where the initial state is already critical and the perturbation couples directly to the order parameter. Such settings are not only important for identifying fundamental organizing principles of nonequilibrium dynamics, but also for present-day experimental platforms, such as quantum simulators and quantum computers, where the corresponding dynamics is directly accessible today.

In this work, we identify a previously unknown form of universal far-from-equilibrium dynamics in Ising systems following a sudden symmetry-breaking field quench at a continuous quantum or thermal phase transition. Such a quench drives the system far beyond linear response and generates superextensive energy fluctuations~\cite{heyl_quenching_2017}.

Our key observation is that the order-parameter fluctuations exhibit a compelling temporal collapse across a wide range of system sizes and quench strengths, pointing to an emergent single-variable scaling form that requires an additional dynamical exponent. We observe this behavior in the 2D quantum Ising model and in the 3D and 4D classical Ising models with Glauber dynamics~\cite{glauber_ising_1963}, whereas it is absent in the 1D quantum and 2D classical cases, providing evidence for a lower critical effective dimension.

A central challenge is the 2D quantum Ising model, where the initial state is itself a nontrivial critical many-body state and the subsequent symmetry-breaking quench generates strongly out-of-equilibrium real-time dynamics. We address this using a Neural Quantum State (NQS) approach~\cite{carleo_solving_2017,schmitt_quantum_2020}, which enables controlled simulations for systems of up to $16\times16=256$ quantum spins.

Our results based on general scaling arguments suggest that the observed universal far-from-equilibrium scaling may extend beyond the Ising cases studied here and could more broadly characterize systems with non-conserved order parameters, opening new avenues for exploring universal dynamics both theoretically and in current experimental platforms.

\paragraph{Protocol.}
We consider a generic protocol for studying far-from-equilibrium dynamics at continuous phase transitions. The system is initially prepared in a state $\rho$ at a continuous phase transition. This is either the ground state $\rho = |\psi_0^{\mathrm{crit}}\rangle\langle \psi_0^{\mathrm{crit}}|$ at a quantum critical point, or the Gibbs state $\rho = \exp(-\beta H^{\mathrm{crit}})/Z$ at a classical thermal phase transition, with $H^{\mathrm{crit}}$ denoting the critical Hamiltonian. At $t=0$, we perform a sudden symmetry-breaking quench by switching on a field $h$ that couples to the order parameter $O$. The post-quench Hamiltonian reads $H = H^{\mathrm{crit}} - h O$, and for times $t>0$ the system evolves either according to the Schr\"odinger equation for quantum systems or according to the corresponding classical dynamics.

Crucially, we consider only cases where the order parameter is not conserved, since otherwise the quench would induce trivial dynamics. The initial state at the continuous phase transition satisfies $\langle O \rangle_0 = 0$, where $\langle \cdot \rangle_0$ denotes the average in the initial state $\rho$. Consequently, $\langle H \rangle_0 = \langle H^{\mathrm{crit}} \rangle_0 = E_0^{\mathrm{crit}}$. However, the energy fluctuations are nonzero and given by
\begin{equation}
\label{eq:energy-variance}
(\Delta E_0)^2 = \langle H^2 \rangle_0 - \langle H \rangle_0^2 = h^2 \langle O^2 \rangle_0 \sim h^2 L^{\kappa},
\end{equation}
with \(\kappa=d+2-z-\eta\) for quantum critical points and \(\kappa=d+2-\eta\) for classical thermal transitions, where $d$ is the spatial dimension, $\eta$ the anomalous dimension, and $z$ is the dynamical critical exponent entering the effective dimension $d+z$.

Importantly, these energy fluctuations are superextensive in system size. As long as $h$ is fixed and independent of system size, this symmetry-breaking quench drives the system beyond linear response, even when $h$ is arbitrarily small.

\paragraph{Model.}
We consider Ising models on periodic hypercubic lattices with $N=L^d$ spins and set $\hbar = k_B = 1$. For the quantum cases $d=1,2$, we study the transverse-field Ising model
\begin{equation}
H^{\mathrm{crit}} = -J \sum_{\langle i,j \rangle} \sigma_i^z \sigma_j^z - g_c \sum_i \sigma_i^x,
\end{equation}
with $J>0$ and nearest-neighbor coupling $\langle i,j \rangle$. The initial state is the critical ground state at $g_c^{\mathrm{1D}}/J = 1$~\cite{pfeuty_one-dimensional_1970} in 1D and at $g_c^{\mathrm{2D}}/J \simeq 3.044$~\cite{bloete_cluster_2002,hesselmann_thermal_2016} in 2D.

For the classical analogue, we focus in the main text on the $d=3$ and $d=4$ Ising models with
\begin{equation}
H^{\mathrm{crit}} = -J \sum_{\langle i,j \rangle} \sigma_i^z \sigma_j^z.
\end{equation}
These systems are initialized at the thermal critical points $T_c^{\mathrm{3D}}/J \simeq 4.5115$~\cite{ferrenberg_pushing_2018} and $T_c^{\mathrm{4D}}/J \simeq 6.6803$~\cite{lundow_revising_2023}, and their dynamics after the quench follows Glauber dynamics~\cite{glauber_ising_1963}.

In all cases the order parameter is the longitudinal magnetization
\begin{equation}
M = \sum_i \sigma_i^z,
\end{equation}
and the quench switches on a uniform longitudinal field, so that
\begin{equation}
H = H^{\mathrm{crit}} - hM.
\end{equation}

\paragraph{Neural Quantum States.}
The 2D quantum case poses the central numerical challenge of this work: real-time quantum dynamics in two dimensions remains difficult for large system sizes, and here the problem is compounded by the fact that the initial state is itself a strongly correlated critical many-body state. We address this using a Neural Quantum State (NQS) approach~\cite{carleo_solving_2017,schmitt_quantum_2020}, in which the many-body wave function is represented in the computational basis as
\begin{equation}
\ket{\psi_{\vec{\theta}}} = \sum_{\vec{s}} \psi_{\vec{\theta}}(\vec{s}) \ket{\vec{s}},
\end{equation}
where $\vec{\theta}$ denotes a set of complex variational parameters. We prepare the critical ground state of $H^{\mathrm{crit}}$ variationally and subsequently evolve the post-quench dynamics in real time using the time-dependent variational principle (TDVP)~\cite{haegeman_tdvp_2011,schmitt_quantum_2020}. This approach gives us access to systems of up to $16\times16=256$ quantum spins, well beyond exact diagonalization, and captures the full sequence from critical-state preparation to long-time nonequilibrium dynamics. In this setting,  a key challenge is that the same variational manifold must accurately describe both the critical initial state and the ensuing far-from-equilibrium evolution. We assess the accuracy of the time evolution through energy conservation and the integrated TDVP residual introduced in Ref.~\cite{schmitt_quantum_2020}; further NQS/TDVP details are provided in the End Matter and Supplemental Material.

\begin{figure}[t]
    \centering
    \includegraphics[width=\columnwidth]{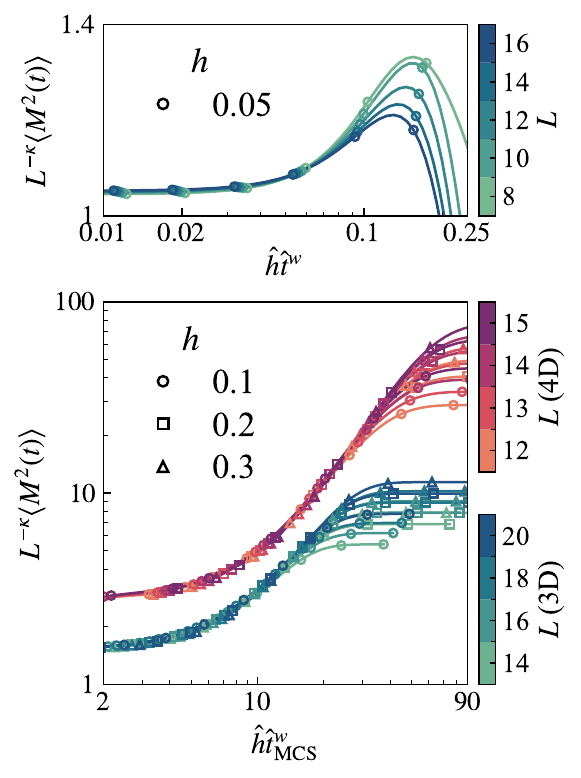}
    \caption{
        \textbf{Dimensional dependence of universal far-from-equilibrium scaling.}
        \textbf{(Top)} Rescaled order-parameter fluctuations $L^{-\kappa}\langle M^2(t) \rangle$ for the 1D TFIM versus $\hat h \hat t^w$, with $\hat t=(Jt)L^{-z}$ and $\hat h=(h/J)L^{y_h}$.
        The apparent scaling regime shrinks with increasing system size $L$, indicating that the collapse is a finite-size artifact rather than true universality.
        \textbf{(Bottom)} Rescaled order-parameter fluctuations $L^{-\kappa}\langle M^2(t) \rangle$ for the 3D and 4D classical Ising models after a quench at $T_c$ (Glauber dynamics), shown versus $\hat h \hat t^w$, with $\hat t=t_{\text{MCS}}L^{-z}$ and $\hat h=(h/J)L^{y_h}$.
        The collapse improves for larger $L$; for the 3D and 4D classical models, the best collapses are obtained for $w=0.853 \pm 0.034$ and $w=0.932 \pm 0.036$, respectively (see Supplemental Material for details).
    }
    \label{fig:no_collapse_1d}
\end{figure}

\paragraph{Results.}
The key result of our work is that the order-parameter fluctuations $\langle M^2(t)\rangle$ exhibit a previously unknown universal dynamical behavior.

In Fig.~\ref{fig:collapse}(c) we show the dynamics of $\langle M^2(t)\rangle$ for different values of the quenched longitudinal field $h$ and for different system sizes up to $N=16\times16=256$ quantum spins. As displayed in Fig.~\ref{fig:collapse}(c)--(e), we first account for the equilibrium finite-size dependence $\langle M^2(t=0)\rangle \propto L^{\kappa}$. The rescaled fluctuations $L^{-\kappa}\langle M^2(t)\rangle$ therefore collapse at short times, as shown in Fig.~\ref{fig:collapse}(d). This short-time collapse is fixed by equilibrium criticality, but at larger times the different curves still deviate substantially.

Our key observation is that this remaining dependence can be absorbed by introducing an additional dynamical exponent $w$: a compelling collapse across all considered field strengths $h$ and system sizes $L$ emerges when the data are plotted as a function of the single variable $\hat h \hat t^w$. Here,
\begin{equation}
\hat t = (Jt)L^{-z}, \qquad \hat h = (h/J)L^{y_h},
\end{equation}
where $z$ is the dynamical critical exponent controlling the finite-size rescaling of time and $y_h$ is the renormalization-group dimension of the longitudinal field $h$ conjugate to $M$. For the quantum Ising models considered here, $z=1$ and $y_h=(d+z+2-\eta)/2$. For Glauber dynamics in the classical Ising models, we use $z \simeq 2.02$ in 3D~\cite{hasenbusch_dynamic_2020} and $z=2$ in 4D~\cite{hohenberg_theory_1977,tauber_critical_2014}, while $y_h=(d+2-\eta)/2$.

For the 2D quantum model, we obtain $w=1.86 \pm 0.11$ from optimizing the data collapse. Importantly, the collapse extends to larger and larger times as the system size increases, indicating that this universal scaling regime becomes more pronounced towards the thermodynamic limit.

We now turn to the generality of this observation. The upshot is that the observed pattern is consistent with this universal dynamics occurring in both quantum and classical systems only above an apparent lower critical effective dimension, see Fig.~\ref{fig:no_collapse_1d}.

In the quantum model, no data collapse is achievable in 1D, see Fig.~\ref{fig:no_collapse_1d}(top). While the short-time behavior, dictated by static equilibrium properties, exhibits a collapse, the behavior at later times is opposite to the 2D case: in 1D the apparent collapse moves to shorter and shorter times as the system size increases. Accordingly, no robust value of $w$ can be assigned in that case.

In the classical models, see Fig.~\ref{fig:no_collapse_1d}(bottom), we observe a compelling data collapse similar to the 2D quantum model also in 3D and 4D. For the 3D and 4D classical models, we obtain $w=0.853 \pm 0.034$ and $w=0.932 \pm 0.036$, respectively. For the 2D classical model, we do not find a global data collapse (see Supplemental Material). At the level of effective dimension, this mirrors the absence of collapse in the 1D quantum case, whereas collapse is present in the 2D quantum and 3D and 4D classical models. This pattern points to a lower critical effective dimension for this far-from-equilibrium scaling phenomenon. We emphasize, however, that this comparison is only heuristic, since in the far-from-equilibrium setting considered here a strict quantum-classical mapping is not expected. Correspondingly, the extracted values of $w$ for the 2D quantum and 3D classical cases are different.

\paragraph{Renormalization group and scaling analysis.}
This collapse can be organized within a dynamic finite-size scaling framework (DFSS)~\cite{hohenberg_theory_1977,tauber_critical_2014,pelissetto_dynamic_2018}.

At criticality, the algebraic decay of equal-time correlations implies that the equilibrium fluctuation of the global order parameter $O=\sum_i O_i$ scales as
\[
\langle O^2\rangle_0 \sim L^{\kappa},
\]
with \(\kappa\) defined above.

Treating $h$ as the relevant scaling field and $L$ and $t$ as the finite-size and dynamical variables, respectively, DFSS predicts the general two-parameter scaling form~\cite{pelissetto_dynamic_2018}
\begin{equation}
\label{eq:dfss-general}
\langle O^2(t;L,h)\rangle = L^{\kappa}\, F(\hat t,\hat h),
\end{equation}
where the reduced variables $\hat t$ and $\hat h$ are the same as those used in Figs.~\ref{fig:collapse} and~\ref{fig:no_collapse_1d}.

The data indicates that, in the observed scaling regime, the two-variable DFSS form effectively collapses onto a single-variable dependence:
\begin{equation}
\label{eq:strongfield_reduction}
F(\hat t,\hat h)\xrightarrow{\quad}\widetilde{\mathcal{F}}(\hat h \hat t^w).
\end{equation}
Within this empirical single-variable reduction, $w$ appears as an additional nonequilibrium dynamical exponent. Once supplemented by the empirically motivated single-variable reduction ansatz, the scaling framework naturally accounts for the observed collapse. The resulting scaling ansatz,
\[
L^{-\kappa}\,\langle M^2(t)\rangle=\widetilde{\mathcal{F}}\!\left(\hat h \hat t^w\right),
\]
captures the observed collapse across different system sizes $L$ and field strengths $h$.

At present, this single-variable scaling form is established only for the order-parameter fluctuations. Whether a similar reduction occurs for other observables remains an open question for the future.

\paragraph{Discussion \& outlook.}
We have identified a previously unknown universal far-from-equilibrium scaling regime that emerges after symmetry-breaking quenches at continuous Ising transitions. Its central signature is a compelling single-variable collapse of the order-parameter fluctuations across system sizes and quench strengths, with the pattern observed across the models studied here being consistent with an apparent lower critical effective dimension, and which within a scaling description requires the introduction of an additional dynamical exponent $w$.

This behavior is distinct from several established forms of universal dynamics. Conventional near-equilibrium relaxation is governed by hydrodynamic modes and linear-response theory~\cite{hohenberg_theory_1977,tauber_critical_2014,pelissetto_dynamic_2018}. By contrast, our protocol generates superextensive energy fluctuations that render linear-response theory inapplicable, placing the system deep in a far-from-equilibrium regime beyond a hydrodynamic description~\cite{heyl_quenching_2017}. It also differs from initial-slip scaling, which concerns the growth of the mean order parameter after a quench from a disordered state with a small symmetry-breaking seed~\cite{janssen_new_1989,zheng_generalized_1996,okano_shorttime_1997,zheng_monte_1998,rossini_coherent_2021,chiocchetta_shorttime_2015,chiocchetta_lightcone_2016}; here, by contrast, the universal scaling identified appears in order-parameter fluctuations rather than in the mean order parameter. Likewise, earlier dynamic finite-size scaling studies of soft quenches with $h\sim L^{-y_h}$ are controlled by known equilibrium exponents~\cite{pelissetto_dynamic_2018,rossini_coherent_2021}, whereas our fixed-field protocol points to an additional exponent $w$.

From a coarse-grained perspective, the classical protocol can be viewed as a Model-A evolution in a Landau-Ginzburg free-energy landscape. This places it near the broader literature on instability-driven phase ordering and spinodal decomposition~\cite{cahn_spinodal_1961,bray_phase_ordering_1994}, but the quench considered here is different: rather than quenching into a negative-curvature region and studying domain coarsening, we start from the critical point and suddenly tilt the order-parameter landscape with a symmetry-breaking field. Developing a coarse-grained theory of this tilted critical dynamics may therefore provide a route toward deriving the exponent $w$ and clarifying its relation, or lack thereof, to conventional phase-ordering growth laws.

Several directions now appear particularly natural. On the experimental side, the quantum protocol should be accessible in programmable Ising-type platforms such as Rydberg arrays and trapped ions~\cite{bernien_rydberg_2017,labuhn_tunable_2016,schauss_crystallization_2015,keesling_quantum_2019,britton_ising_2012,monroe_trapped_ions_2021}. On the theoretical side, it will be important to determine whether this scaling extends beyond Ising systems with non-conserved order parameters, and whether related single-variable behavior appears in observables other than the order-parameter fluctuations. Most importantly, the microscopic origin of the exponent $w$ remains open. Clarifying this point will be essential for understanding whether the scaling identified here reflects a broader organizing principle of far-from-equilibrium critical dynamics. More broadly, our findings suggest that universal scaling far from equilibrium may be more common than previously recognized, opening new avenues for theoretical investigation and experimental verification in quantum and classical many-body systems.

\paragraph{Acknowledgments.}
We also acknowledge fruitful discussions with A. Scardicchio, C. A. Weber, and A. Sharma.
The NQS dynamics is simulated using the \texttt{jVMC} package~\cite{Schmitt2022}.
This project has received funding from the European Research Council (ERC) under the European Union's Horizon 2020 research and innovation programme (Grant Agreement No.~853443).
 This work was supported by the German Research Foundation (DFG) via project 492547816 (TRR~360).
 We gratefully acknowledge the scientific support and high-performance computing resources provided by the LiCCA HPC cluster at the University of Augsburg and by the Jülich Booster (JUWELS Booster) at the Jülich Supercomputing Centre, Forschungszentrum Jülich (project \texttt{nqscm}). The LiCCA cluster is co-funded by the DFG (Project ID 499211671).

\noindent T.W. performed the calculations. T.W., L.B., and M.H. discussed and interpreted the results and wrote the manuscript.

\noindent The authors declare no competing interests.

\paragraph{Data availability\textemdash} The data contained in all figures of this article is available on Zenodo~\cite{Zenodo}.

\bibliography{sources}

\appendix*
\setcounter{equation}{0}
\renewcommand{\theequation}{A\arabic{equation}}
\section{Neural Quantum State simulations}
\label{sec:endmatter-nqs}
To obtain the 2D quantum data in Figs.~\ref{fig:collapse} and~\ref{fig:no_collapse_1d}, we use the Neural Quantum States (NQS) framework~\cite{carleo_solving_2017,czischek_quenches_2018,schmitt_classical_2018,fabiani_ultrafast_2019,wu_ann_otoc_2020,choo_two_dimensional_2019}. The many-body wave function is represented in the computational basis as
\begin{equation}
\ket{\psi_{\vec{\theta}}} = \sum_{\vec{s}} \psi_{\vec{\theta}}(\vec{s}) \ket{\vec{s}},
\end{equation}
where $\vec{\theta}$ denotes a set of complex variational parameters. In practice, we employ a translation-equivariant complex-valued convolutional ansatz; explicit architectural details are given in the Supplemental Material. Because the longitudinal field explicitly breaks the $\mathbb{Z}_2$ spin-flip symmetry, we do not constrain the variational manifold to the symmetric sector.

The initial critical ground state of $H^{\mathrm{crit}}$ is prepared variationally using stochastic reconfiguration with Monte Carlo sampling. Starting from this optimized state, the post-quench unitary dynamics under $H=H^{\mathrm{crit}}-hM$ are computed using the time-dependent variational principle (TDVP)~\cite{haegeman_tdvp_2011,carleo_solving_2017,schmitt_quantum_2020}, which yields the effective equation of motion
\begin{equation}
S_{k,k'} \dot{\theta}_{k'} = -i F_k .
\end{equation}
This setting is particularly demanding because the same variational manifold must accurately capture both the critical initial state and the ensuing far-from-equilibrium dynamics.

To assess the reliability of the simulations, we monitor conservation of $\langle H\rangle$ and the integrated TDVP residual introduced in Ref.~\cite{schmitt_quantum_2020}, and we verify convergence with respect to the number of channels in the ansatz. Further architectural details, sampling parameters, representative hyperparameters, and convergence data are provided in the Supplemental Material.

\beginsupplement

\begin{center}
\textbf{\large Supplemental Material: Universal Dynamics for Symmetry-Breaking Quenches at Continuous Phase Transitions: Evidence for a New Dynamical Exponent}
\end{center}

\renewcommand{\thesection}{\Roman{section}}
\renewcommand{\thesubsection}{\Alph{subsection}}

\makeatletter
\let\orig@section\section
\renewcommand{\section}[1]{%
  \refstepcounter{section}%
  \orig@section*{\thesection.\quad #1}%
}%
\let\orig@subsection\subsection
\renewcommand{\subsection}[1]{%
  \refstepcounter{subsection}%
  \orig@subsection*{\thesubsection.\quad #1}%
}%
\makeatother

\noindent
Section~\ref{sec:S_nqs} summarizes the Neural Quantum State (NQS) ansatz and TDVP accuracy diagnostics underlying the quantum data in Figs.~\ref{fig:collapse} and~\ref{fig:no_collapse_1d}.
Section~\ref{sec:2d-classic-crossing} documents the single-crossing behavior referenced in the main text for the 2D classical model.
Section~\ref{sec:S_collapse} details the collapse optimization used to determine the reported values of $w$.
Equations and figures introduced in this supplement carry an ``S'' prefix; unprefixed references point to the main paper.
\vspace{0.75em}
\section{Neural Quantum State simulations}\label{sec:S_nqs}

\noindent
This section provides the essential NQS/TDVP details needed to reproduce and assess the quantum quench data shown in the main text, including the convolutional ansatz, optimization workflow, and quantitative accuracy diagnostics.

\subsection{Ansatz}\label{sec:S_ansatz}
We use a translation-equivariant complex-valued convolutional ansatz with $C$
channels. For a spin configuration $s$, the pre-activation is
\begin{equation}
z_{c,k}(s)=\sum_{i,j=1}^{F} W_{c,i,j}\, s_{T_k(i,j)} + b_c,
\end{equation}
where $T_k$ denotes the action of the 2D lattice translation group $\mathbb{Z}_L \times \mathbb{Z}_L$ on the filter coordinates, shifting to all $N=L^2$ positions with periodic boundary conditions.
The activation is $a_{c,k}=f(z_{c,k})$ with
$f(z)=z-\frac{z^3}{3}+\frac{2}{15}z^5$.
We set $F=L$ to capture critical long-range correlations.
The log-amplitude is
\begin{equation}
\ln\psi_{\theta}(s)=\frac{1}{\sqrt{CN}}\sum_{c=1}^{C}\sum_{k=1}^{N} a_{c,k}(s),
\qquad N=L^2.
\end{equation}
The number of real parameters is $N_p=2C(N+1)$.

\subsection{Optimization and time evolution}\label{sec:S_tdvp}
The critical ground state is prepared variationally, and real-time dynamics are
computed using TDVP. We monitor energy conservation and the TDVP residual
(Sec.~\ref{sec:S_tdvp_residual}) as accuracy diagnostics.
Table~\ref{tab:hyperparameters} lists representative hyperparameters.

\begin{table}[htbp]
\centering
\begin{tabular}{ll}
\hline
\textbf{Markov Chain (Ground State Preparation)}           & \\
\hline
Number of Samples               & 5000 \\
Sweep Steps                     & N \\
Thermalization Sweeps          & 1 \\
Number of Chains    & 500 \\
\hline
\textbf{Markov Chain (Real Time Dynamics)}           & \\
\hline
Number of Samples              & $4 \times 10^5$ \\
Sweep Steps                    & $3N$ \\
Thermalization Sweeps          & 3 \\
Number of Chains (per gpu)     & 500 \\
\hline
\textbf{TDVP(Ground State Preparation)}                   & \\
\hline
Initial Diagonal Shift          & 10 \\
Shift Interval                  & 50 \\
Shift Factor                    & $10^{-1}$ \\
Minimum Diagonal Shift          & $10^{-10}$ \\
Initialization Steps           & $5000$ \\
Euler Time Step                 & $10^{-2}$ \\
\hline
\textbf{TDVP(Real Time Dynamics)}                   & \\
\hline
Integrator tolerance            & $10^{-5}$ \\
snrTol                          & $2$ \\
pinvTol                         & $10^{-7}$ \\
\hline
\textbf{CpxCNN}                 & \\
\hline
Filter Size                     & $L\times L$ \\
Channels                        & 10 \\
Bias                            & True \\
Activation Function              & Poly5 \\
\hline
\end{tabular}
\caption{Representative hyperparameters used for ground-state preparation and real-time TDVP evolution.}
\label{tab:hyperparameters}
\end{table}

\subsection{TDVP residual}\label{sec:S_tdvp_residual}
We use the dimensionless TDVP residual~\cite{schmitt_quantum_2020}
\begin{equation}
r^2(t)=
\frac{D^2\!\bigl(\ket{\psi_{\theta(t+\delta t)}}, e^{-iH\delta t}\ket{\psi_{\theta(t)}}
\bigr)}
     {D^2\!\bigl(\ket{\psi_{\theta(t)}}, e^{-iH\delta t}\ket{\psi_{\theta(t)}}
\bigr)},
\end{equation}
with $D$ the Fubini--Study distance. As a cumulative measure we report
\begin{equation}
R^2(t)=\int_0^t dt'\, r^2(t'),
\end{equation}
evaluated as a discrete sum over time steps.

\subsection{Convergence analysis}
\label{sec:convergence}

To ensure the reliability of our Neural Quantum State simulations, we systematically assess convergence with respect to the number of channels $C$ in the convolutional ansatz, addressing (i) the accuracy of the initial critical ground state preparation, and (ii) the stability and accuracy of the subsequent real-time dynamics.
The convergence checks shown here were performed for $L=10$ and $h=0.1$.

\begin{figure}[t]
    \centering
    \includegraphics[width=\columnwidth]{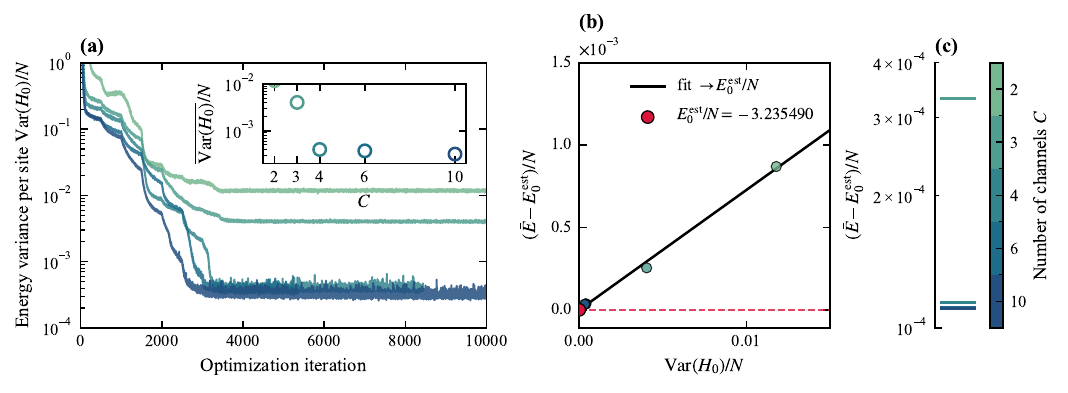}
    \caption{
        \textbf{Ground-state convergence analysis.}
        (a) Energy variance per site $\mathrm{Var}(H^{\mathrm{crit}})/N$ versus optimization iteration; inset: plateau-averaged $\overline{\mathrm{Var}(H^{\mathrm{crit}})}/N$ versus $C$.
        (b) Energy-variance extrapolation: $(\bar{E} - E_0^{\mathrm{est}})/N$ versus $\mathrm{Var}(H^{\mathrm{crit}})/N$; the linear fit yields $E_0^{\mathrm{est}}$.
        (c) Energy deviation $(\bar{E} - E_0^{\mathrm{est}})/N$ for each $C$ (horizontal segments).
        All quantities show systematic convergence with increasing $C$.
        Data shown for $L=10$ and $h=0.1$.
    }
    \label{fig:convergence-groundstate}
\end{figure}

Figure~\ref{fig:convergence-groundstate} shows the convergence of the initial ground-state preparation with respect to $C$.
Panel (a) displays the energy variance $\mathrm{Var}(H^{\mathrm{crit}})/N$ during optimization (inset: plateau-averaged variance versus $C$).
Panel (b) shows the energy-variance extrapolation used to estimate the ground-state energy $E_0^{\mathrm{est}}$.
Panel (c) summarizes the energy deviation $(\bar{E} - E_0^{\mathrm{est}})/N$ for each $C$.
All quantities show systematic convergence with increasing $C$.

\begin{figure}[t]
    \centering
    \includegraphics[width=\columnwidth]{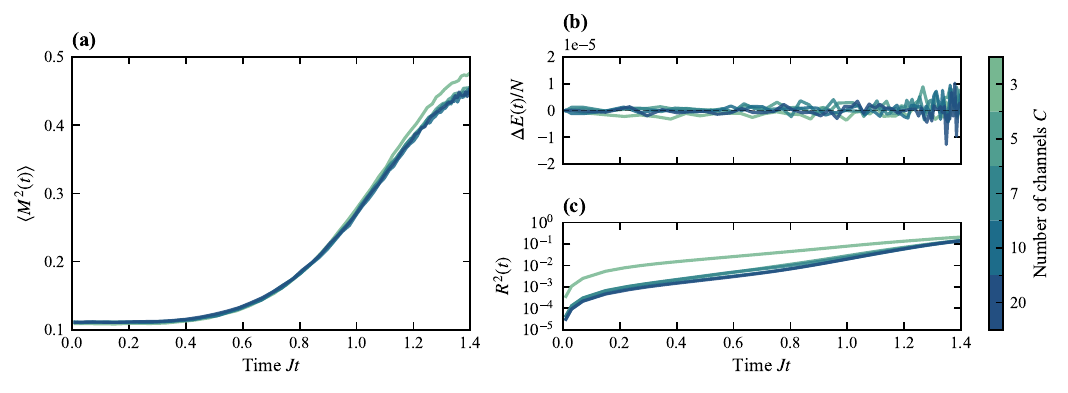}
    \caption{
        \textbf{Dynamics convergence analysis.}
        Convergence of the post-quench dynamics with respect to $C$.
        (a) Order-parameter fluctuations $\langle M^2(t)\rangle$ for different $C$.
        (b) Energy conservation $\Delta E(t)/N = (\langle H\rangle(t) - \langle H\rangle(0))/N$.
        (c) Integrated TDVP residual $R^2(t)$.
        The dynamics remain stable and accurate across the full time window.
        Data shown for $L=10$ and $h=0.1$.
    }
    \label{fig:convergence-dynamics}
\end{figure}

Figure~\ref{fig:convergence-dynamics} demonstrates the convergence of the real-time dynamics following the quench.
The order-parameter fluctuations $\langle M^2(t)\rangle$ converge with respect to $C$, energy conservation remains excellent throughout the evolution, and the integrated TDVP residual $R^2(t)$ stays small, confirming that the time evolution is accurate and stable.

\section{Single-crossing behavior in the 2D classical model}
\label{sec:2d-classic-crossing}

\noindent
In the main text we note that, unlike in $d=3,4$, the 2D classical Ising model does not exhibit a global single-variable collapse; instead, the rescaled curves for different $L$ intersect only locally.
Here we show this ``single crossing'' explicitly and quantify it via the relative spread across sizes.

\begin{figure}[ht]
    \centering
    \includegraphics[width=\columnwidth]{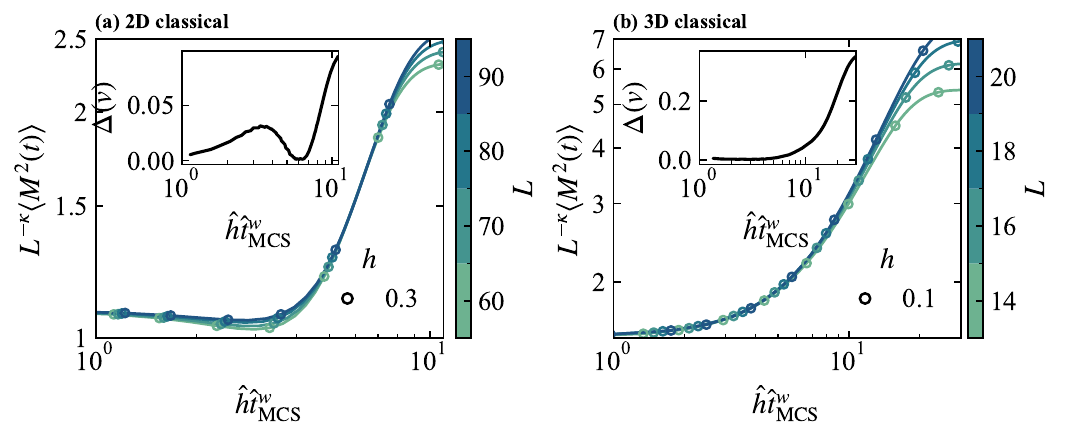}
    \caption{
        \textbf{Single-crossing behavior in the rescaled observable (2D and 3D classical).}
        Left: 2D classical Ising---$L^{-\kappa}\langle M^2(t)\rangle$ versus $\hat h \hat t^w$ for $L=\{60,70,80,90\}$ at $h=0.3$, using the illustrative plotting value $w=0.775$, where $\hat t=t_{\text{MCS}}L^{-z}$, $\hat h=(h/J)L^{y_h}$, and $\kappa=d+2-\eta$ for classical models. Right: 3D classical Ising---same rescaling for $L=\{14,16,18,20\}$ at $h=0.1$, using the reported value $w=0.853$. Time is measured in Monte Carlo sweeps $t_{\text{MCS}}$. Insets: relative spread $\Delta(\hat h \hat t^w)$ across $L$; the sharp minimum demonstrates that the curves coincide only locally (a single crossing) rather than exhibiting a global collapse. The left-panel value $w=0.775$ is introduced only to visualize the local crossing and is not interpreted as a reported collapse exponent.
    }
    \label{fig:2d_classic_crossing}
\end{figure}

Figure~\ref{fig:2d_classic_crossing} shows the finite-size scaling plot of the rescaled observable $\widetilde{M}^2_L(t)\equiv L^{-\kappa}\langle M^2(t)\rangle$ with $\kappa=d+2-\eta$ for classical models, versus the rescaled time $\hat h \hat t^w$, where $\hat t=t_{\text{MCS}}L^{-z}$ and $\hat h=(h/J)L^{y_h}$, and $t_{\text{MCS}}$ is measured in Monte Carlo sweeps (not $Jt$). For the 2D classical panel we use the illustrative plotting value $w=0.775$ only to display the local crossing; since no global collapse is obtained, we do not assign a reported value of $w$ to that model. While the curves for different $L$ intersect at a common point, they do \emph{not} exhibit a global data collapse: away from the intersection the curves separate systematically with $L$.

To quantify that this is a \emph{single crossing} rather than a collapse, we compute at fixed $\hat h \hat t^w$ the \textbf{relative spread across sizes} $\Delta(\hat h \hat t^w)$,
\begin{equation}
    \Delta(\hat h \hat t^w) \equiv
    \frac{\max_L \widetilde{M}^2_L(\hat h \hat t^w)-\min_L \widetilde{M}^2_L(\hat h \hat t^w)}{\overline{\widetilde{M}^2}(\hat h \hat t^w)},\qquad
    \overline{\widetilde{M}^2}(\hat h \hat t^w)=\frac{1}{N_L}\sum_L \widetilde{M}^2_L(\hat h \hat t^w),
\end{equation}
where the max/min/mean are taken over the set of system sizes shown. The inset displays a pronounced minimum, coincident with the apparent intersection, making explicit that the agreement across $L$ is localized rather than reflecting a full collapse.

\section{Collapse procedure}\label{sec:S_collapse}

\noindent
This section summarizes how we determine the reported values of the exponent $w$ appearing in the single-variable scaling form discussed in the main text. The goal is to choose $w$ such that, after the known finite-size rescaling, data taken at different $(L,h)$ fall onto a single curve when plotted versus the combined variable $\hat h\,\hat t^{\,w}$.

\paragraph{Rescaling.}
For each dataset labeled by $(L,h)$ we form the rescaled observable
\begin{equation}
\widetilde M^2(t;L,h)=L^{-\kappa}\langle M^2(t;L,h)\rangle,
\qquad
\kappa=
\begin{cases}
d+2-z-\eta, & \text{quantum},\\
d+2-\eta, & \text{classical},
\end{cases}
\end{equation}
and plot it against the collapse variable
\begin{equation}
x=\hat h\,\hat t^{\,w},\qquad
\hat h=(h/J)L^{y_h},
\end{equation}
with
\begin{equation}
\hat t=(Jt)L^{-z}\quad \text{for quantum models},\qquad
\hat t=t_{\rm MCS}L^{-z}\quad \text{for classical models}.
\end{equation}

\paragraph{Time window.}
At late times, finite-size effects (and, in the quantum case, recurrences) spoil scaling. We therefore restrict the analysis to times before this regime. Concretely, for each curve we define a crossover time $t_x(L,h)$ that marks where $\langle M^2(t)\rangle$ bends away from its early-time behavior in a log--log plot, and we fit only within a relative window
\begin{equation}
t\in[\beta\,t_x(L,h),\,\gamma\,t_x(L,h)],\qquad \beta<\gamma.
\end{equation}
To avoid relying on a single arbitrary choice, we repeat the full analysis over many pairs $(\beta,\gamma)$.

\paragraph{Collapse optimization.}
For a given window $(\beta,\gamma)$ and trial value of $w$, we quantify the quality of the collapse by the curve-to-curve spread of $\log\widetilde M^2$ at common values of $\log x$, restricted to the range of $x$ where the different curves overlap. Concretely, we evaluate each curve $\alpha\equiv(L,h)$ on a common log-spaced grid $\{x_j\}$ within the overlap region (using interpolation in $\log x$), and define
\begin{equation}\label{eq:S_cost_def}
    C(w)=\frac{1}{N_{\rm eff}}\sum_{j\in\mathcal{J}}\mathrm{Var}_{\alpha}\!\left[\log\widetilde M^2_{\alpha}(x_j)\right],
    \qquad
    w_{\rm opt}(\beta,\gamma)=\arg\min_w C(w),
\end{equation}
where the variance is taken over all curves that contribute at $x_j$, $\mathcal{J}$ denotes the set of grid points with sufficient multi-curve coverage, and $N_{\rm eff}=|\mathcal{J}|$.

\paragraph{Final estimate and uncertainty (``$k\sigma$'' band).}
Not all windows yield a meaningful comparison (e.g., if the overlap in $x$ is too small or too few curves contribute), so we keep only windows that pass basic overlap/coverage checks. This produces an ensemble of acceptable window-by-window estimates $\{w_{\rm opt}(\beta,\gamma)\}$. We take the median as the reported value,
\begin{equation}
w_{\rm rep}=\mathrm{median}\!\bigl(w_{\rm opt}(\beta,\gamma)\bigr),
\end{equation}
and define a systematic uncertainty from the central quantile width of the accepted-window distribution,
\begin{equation}
\sigma_{\rm sys}=\tfrac{1}{2}\bigl(q_{84}-q_{16}\bigr),
\end{equation}
where $q_p$ is the $p$th percentile of $\{w_{\rm opt}(\beta,\gamma)\}$. We report a conservative interval $w_{\rm rep}\pm k\,\sigma_{\rm sys}$ (with $k=4$ for the error bars quoted in the main text).

\end{document}